\documentclass[11pt,preprint]{aastex}
\usepackage{geometry} 
\usepackage{amssymb}
\usepackage{epsfig}
\shorttitle{Radio Supernovae}
\shortauthors{Kamble et al.} 

\begin{document}
\title{Radio Supernovae in the Local Universe}
\author{Atish Kamble, Alicia Soderberg, Edo Berger, Ashley Zauderer, Sayan Chakraborti and
	Peter Williams on behalf of a larger collaboration}
\affil{Harvard-Smithsonian Center for Astrophysics,\\
	60 Garden St., Cambridge, MA 02138}
\email{akamble@cfa.harvard.edu}

\begin{abstract}
	In the last three decades, about 50 radio supernovae have been detected as a result 
	of targeted searches of optically discovered supernovae in the local universe. 
	Despite this relatively small number some diversity among them has already been 
	identified which is an indication of the underlying richness of radio supernovae
	waiting to be discovered. For example, 
	comparison of star formation and supernova discovery rate imply that as many as 
	half of the supernovae remain undetected in the traditional optical 
	searches, either because of intrinsic dimness or due to dust obscuration. This has 
	far reaching consequences to the models of stellar and galaxy evolution.
	A radio sky survey would be ideal to uncover larger supernova population. 
	Transient radio sky would benefit significantly from such a survey. 
	With the advent of advanced gravitational wave detectors a new window is set to open 
	on the local Universe. Localization of these gravitational detectors is poor to identify
	electromagnetic counterparts of the gravitational wave sources. However, the longer 
	lasting radio emission accompanied in these sources could be effectively identified 
	in a radio sky survey. We advocate a medium area ($\approx$ a few thousand 
	$\rm deg^{2}$) radio sky survey at C-band. Alternatively, a survey at S-band
	has advantage of larger sky coverage without serious loss of the science case 
	as presented here. Understanding the background in radio sky will be of paramount
	importance for the upcoming sensitive radio facilities including the Square Kilometer Array.
\end{abstract}

\section{Introduction}
	A full characterization of the transient sky is one of the key observational frontiers in modern 
	astronomy as underscored by the Astro2010 Decadal Survey. It is driven by the fact that 
	transient phenomena generally result from a rapid and violent release of energy and 
	therefore provide a unique window into extreme physics, massive star outbursts and 
	deaths, and the diverse behavior of compact objects (white dwarfs, neutron stars, 
	and black holes). Very Large Array (VLA) radio telescope has been a pioneer in the 
	transient astronomy over decades especially through its targeted observations of 
	novae, supernovae (SNe), gamma ray bursts (GRBs) and, recently, 
	tidal disruption events (TDEs). Moreover, it has enabled 
	a systematic characterization of the radio properties of SNe. Detailed comparison of radio 
	observations with optical and other wavebands has revealed a rich diversity among SNe
	such as a broad luminosity distribution, SNe powered by a central engine and some without 
	optical or high energy counterparts to state a few. A radio sky survey would be ideal 
	to address some of
	these issues and to uncover and systematically study a larger supernova population.

	With the advent of synoptic sky surveys astronomy is on the brink of revolution and radio 
	astronomy is finally gearing up for that era, which had until now been largely the domain 
	of optical and other high-frequency bands.
	Several upcoming and operational radio facilities, such as ASKAP-VAST
	\footnote{http://www.physics.usyd.edu.au/sifa/vast/index.php}, 
	MWA \footnote{http://www.mwatelescope.org/}, 
	GMRT-TGSS \footnote{http://tgss.ncra.tifr.res.in/}, LOFAR \footnote{http://www.lofar.org/} etc., 
	are well positioned to do the low frequency ($\leq$ 1.5 GHz) whole sky survey
	and will map the radio sky at those frequencies at regular intervals. VLA should, therefore,
	concentrate on its unique strength at high-frequencies (around 5 GHz) with an eye on 
	future facilities (SKA, \footnote{https://www.skatelescope.org/}, 
	LSST \footnote{http://www.lsst.org/lsst/}, 
	Advanced-LIGO \footnote{https://www.advancedligo.mit.edu/}, 
	Zwicky Transient Facility etc.) and the skies that they open up 
	(novae, SNe, GRBs, TDEs and other transients).

\section{Radio Emission from Supernovae}
	On physical grounds SNe could be broadly classified in two branches: 
	core collapse (type Ib, Ic and type II) and thermonuclear (type Ia). The gravitational
	collapse of stellar core is accompanied by the ejection of outer stellar shells at very 
	high velocities ($\approx 10,000$ km/s) which upon interaction with the 
	circum-stellar material launches a shock-wave propagating at a fraction of 
	the speed of light ($v \approx$ 0.1-0.2 c). The material swept by the shock gets heated
	to high temperatures. Relativistic electrons in this plasma gyrate in the shock 
	amplified magnetic field generating synchrotron radiation which peaks in the radio
	frequency bands. Largely similar description would be applicable  
	in the case of thermonuclear SNe although to-date no radio emission has been seen 
	from the site of type-Ia SNe. Environmental reasons could be strongly to blame
	for those non-detections.
	
	The theory of radio emission due to SN shock wave has been developed 
	by \citet{Chevalier1982}. A significant
	progress in observations has been achieved due to the work of several groups, 
	notably by \citet{Weiler2002, Berger2003, Soderberg2007, Stockdale2006}.

	The synchrotron radio luminosity of a SN could be described in a semi-empirical 
	form
	\begin{equation}
		f_{\nu}(\nu,t) = K_{1} \left( \frac{\nu}{\rm 5 GHz} \right)^{\beta_{f}} 
					\left( \frac{t}{\rm day} \right)^{\alpha_{f}} e^{-\tau_{\nu}}
	\end{equation}
	with the optical depth due to the external absorption
	\begin{equation}
		\tau_{\nu}(t) = K_{2} \left( \frac{\nu}{\rm 5 GHz} \right)^{\beta_{\tau}} 
					\left( \frac{t}{\rm day} \right)^{\alpha_{\tau}}
	\end{equation}
	where $\beta_{\tau}$ and $\beta_{f}$ describe spectral evolution of flux and 
	optical depth,
	respectively. Similarly, $\alpha_{\tau}$ and $\alpha_{f}$ describe temporal evolution of 
	flux and optical depth, respectively.
	
	In addition to the external source of absorption, radiating electrons act as a source 
	of internal absorption via the well known process of synchrotron self-absorption. 
	This prominent spectral feature cascades down in frequencies with time.
	Its detection and evolution is crucial for estimating density of circum-stellar material 
	and the shock speed. Along with the estimation of peak brightness these observed
	parameters are used to infer shock wave kinetic energy.

\begin{figure}	
		\begin{center}
			\includegraphics[width=6cm]{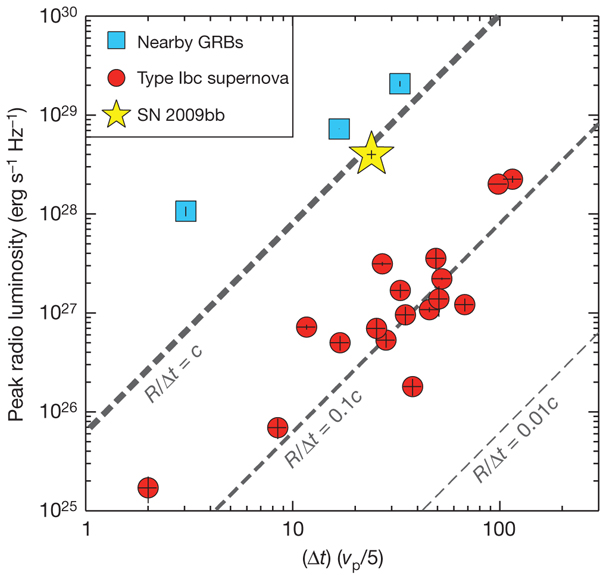}
			\includegraphics[width=6cm]{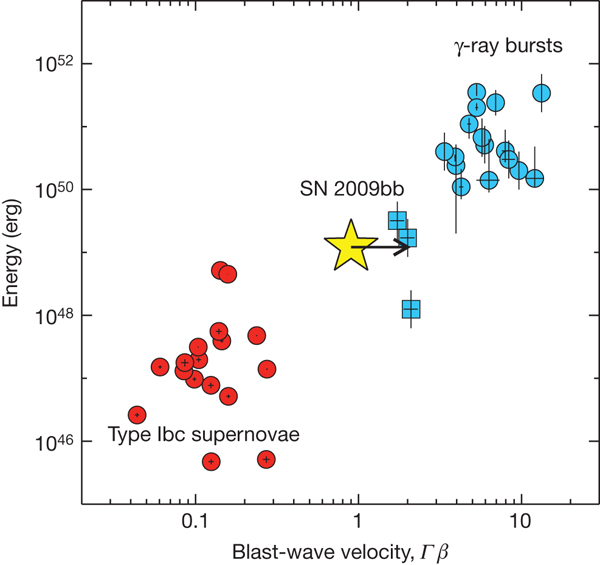}
			\caption{\emph{Left Panel} The peak spectral luminosity at 5 GHz is plotted against 
			the epoch of peak brightness. These observed quantities are related to 
			the shock speed racing down the circum-stellar medium through 
			equation \ref{eqn:Lptp}. The dashed lines corresponds different shock 
			speeds ($\beta_{sh} = 0.01, 0.1, 1$). As can be seen a large population of 
			radio supernovae (red points) drive sub-relativistic shock wave while
			Gamma-ray burst, which are powered by central engine lead to bright 
			radio afterglows due to relativistic shock waves. Engine driven supernovae,
			such as SN 2009bb with mildly relativistic shock wave, occupy the large gap 
			in the middle. \emph{Right Panel} Kinetic energy in the fastest moving ejecta
			is plotted against shock wave speeds and compared for SNe and GRBs.
			It can be seen that the normal supernova shock wave carries a few orders 
			of magnitude less energy than a GRB shock wave. Similar to the velocity space, 
			energy distribution of SN and GRBs also appear bimodal. Events similar to 
			SN 2009bb populating this parameter space remain to be discovered.
			These figures have been taken from \citet{Soderberg2010}.}
			\label{fig:LuminosityVelocity}
		\end{center}
	\end{figure}

	Figure \ref{fig:LuminosityVelocity} is an effective way of looking at the diversity in the 
	population of RSNe. The peak radio spectral luminosity ($L_{\nu,p}$) at 5 GHz and 
	its epoch ($t_{p}$) are related to the shock velocity ($\beta_{sh}$) expressed in the units
	of speed of light \citep{Chevalier1998}. 
	\begin{equation}
		L_{\nu,p} = 1.2 \times 10^{28} \beta_{sh}^{36/17} \left( \frac{\nu_{p}}{\rm 5 ~GHz} \right)^{36/17} 		\left( \frac{t_{p}}{\rm 10 ~day} \right)^{36/17}
		\label{eqn:Lptp}	
	\end{equation}
	
	Supernovae are formed during the collapse of massive star, and merely 1\% 
	($= 10^{51}$ erg) of the total gravitational energy released during the collapse 
	is coupled to the SN ejecta moving at $\approx 10,000$ km/s. Even smaller fraction
	$\lesssim 10^{-5}$ of the total energy is coupled to the fastest moving shock wave 
	responsible for synchrotron emission shining in radio. Although formed under similar 
	circumstances, GRBs are much rarer events. In contrast
	to SNe, shock waves of GRBs are highly relativistic and orders of magnitude more 
	energetic. According to the widely accepted scenario, this difference stems from 
	the way SNe and GRBs derive their energy. SNe derive their energies from 
	the asymmetric core collapse and is coupled to the massive stellar ejecta. GRBs,
	on the other hand, derive most of their energy from the central engine formed
	during the core collapse and drives relativistic jet. Appearance of 
	SN\,1998bw/GRB\,980425 with bright radio emission and relativistic ejecta
	in the intermediate velocity-energy parameter space implies that stellar collapse
	can produce diverse explosions. Subsequent discovery of mildly relativistic SN\,2009bb,
	possibly driven by central engine but  
	without a gamma ray trigger, strongly supports this inferred diversity and 
	has deep significance for the theories of stellar evolution and explosion. 
		
\begin{figure}	
	\begin{center}
		\includegraphics[width=7cm]{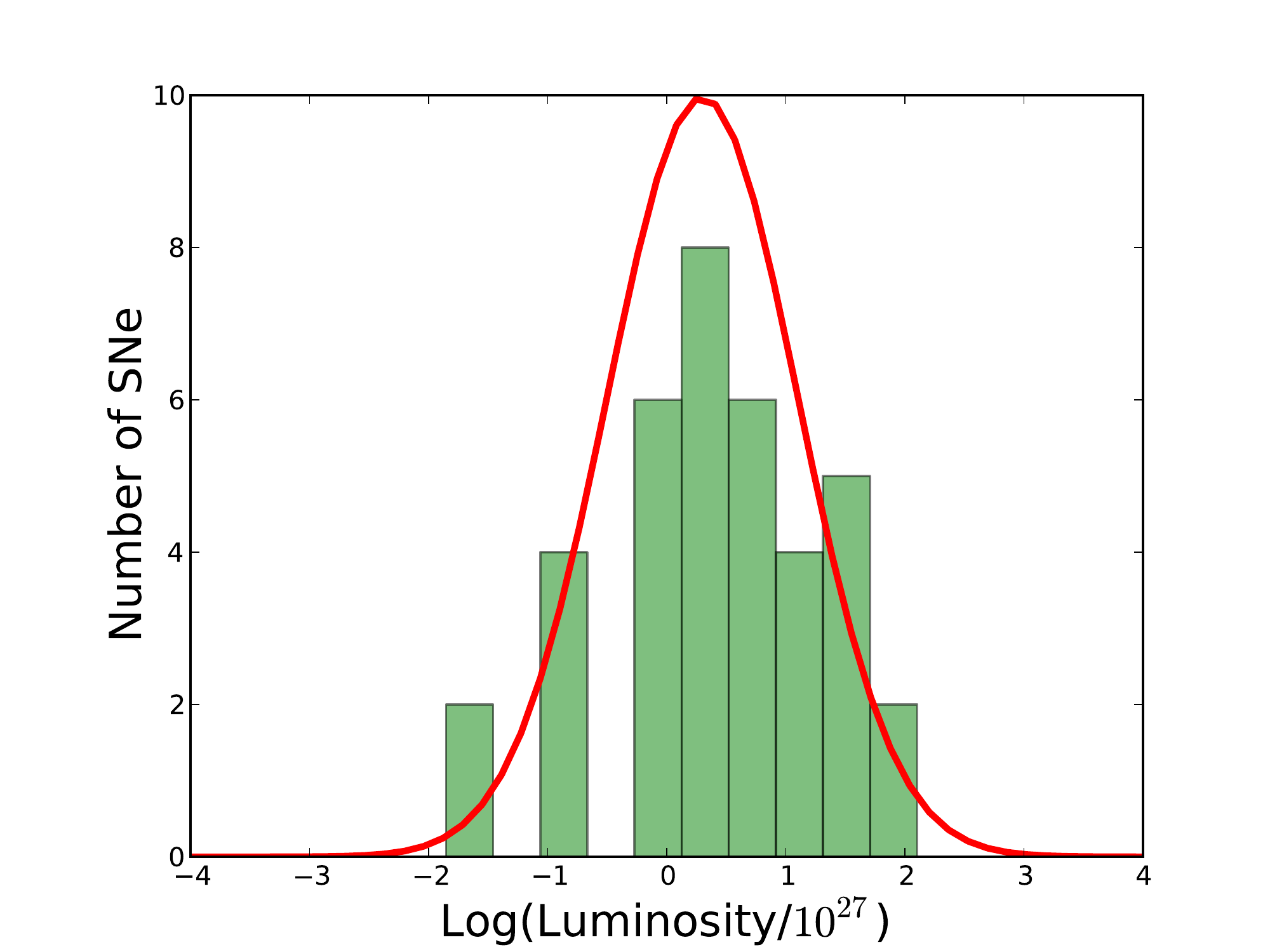}
		\includegraphics[width=7cm]{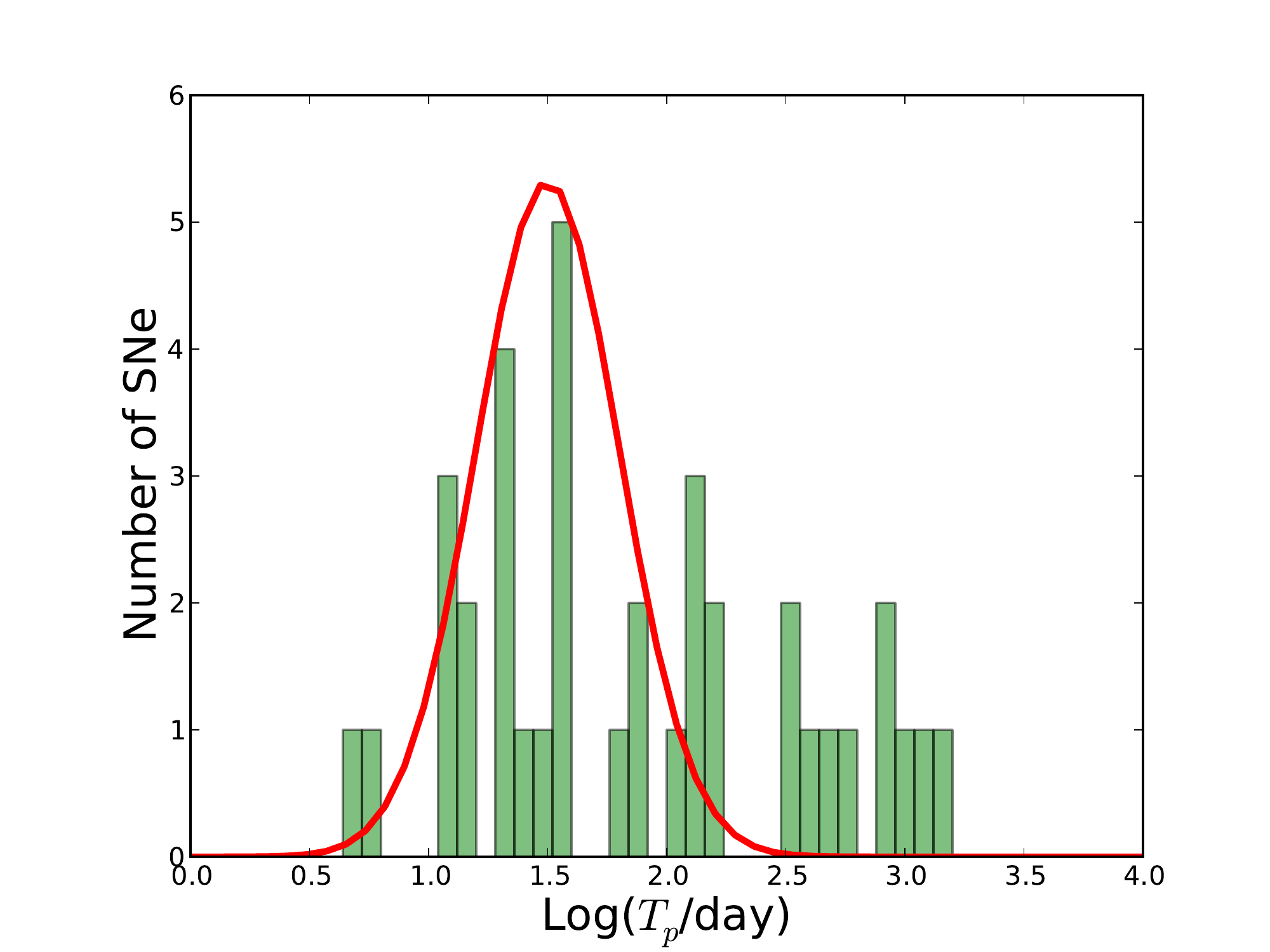}
		\caption{Peak radio luminosity distribution of core collapse SNe is shown in the
		 \emph{left panel}. The mean of the log-normal distribution is around 
		 $log_{10}(L/10^{27}) = 0.3$ 
		and $\sigma = 0.8$. Distribution of the epochs of peak brightness in the 
		\emph{right panel} 
		shows significantly delayed peaks, some of them several hundred days after 
		the supernova. Most of the delayed peaks, such as in the case of SN 1996cr and 
		2001em for example, are thought to be due to the interaction of shock wave with
		dense circum-stellar medium or shell. The red curves have been plotted to guide the eye. 
		The mean of the log-normal curve for the peak times is $T_{p} = 31.6$ days 
		with the standard deviation of 2 days.}
		\label{fig:dist}
	\end{center}
\end{figure}

	As a result of several targeted searches and subsequent monitoring campaigns 
	key properties of radio emission from 
	SNe have been established which will be helpful in guiding future observations. 
	Figure \ref{fig:dist} summarizes two of the most important observational 
	properties of RSNe i.e. observed distributions of peak luminosities and their epochs. 
	Core-collapse RSNe have peak luminosities of $L_{\nu,p} = 2\times 10^{27}$ 
	erg/s/Hz and they attend those peak luminosities around $T_{p} = 31.6 \pm 1.9$ days
	after the SN.
	
\section{Motivation}
	The targeted observations have significantly improved our understanding of the properties 
	of RSNe. At the same time, they have revealed some deep gaps in our knowledge about them.
	For example, what are the progenitors of type Ia SNe ? Where are the off-axis afterglows 
	of GRBs and what's the true census of engine driven SNe ? Below we identify some important 
	gaps and key questions in ours understanding of radio supernovae which also form the 
	motivation for proposing this VLA sky survey.
	
	\begin{enumerate}
		\item Relativistic SNe: The association of SN 1998bw with GRB 980425 brought
		about a significant advancement in our understanding of these relativistic 
		transients. On one side this association was the first proof that long 
		duration GRBs originate 
		in the collapse of massive stars, which was conclusively established subsequently by
		GRB030329/SN2003dh. On the other hand, it opened up a possibility that  
		due to the collimated nature of GRBs, although many such triggers could 
		be missed,
		associated SNe which are more isotropic could be detected. Such SNe powered
		by the ``central engine" are considered the missing link in the core collapse scenario.
		Figure \ref{fig:LuminosityVelocity} shows this picture clearly. GRBs are relativistic 
		($\Gamma \beta \sim$ several) and have energies $\approx 10^{50}-10^{52}$ erg. 
		Normal RSNe have lower energy budgets $\approx 10^{46}-10^{48}$ erg 
		and non-relativistic blast waves ($\beta \sim 0.1$). It was expected that SNe
		with intermediate energies and mildly relativistic velocities, similar to some SN 
		associated with low energy GRBs, should be detectable without associated GRB.
		SN 2009bb with energy $\sim 10^{49}$ erg and $\beta \sim 0.9$ just turned out 
		to be such an object without a gamma ray trigger. Naturally, it is expected that several
		such relativistic SNe should be out there waiting to be discovered.
		
		\item Type Ia supernovae: Despite several deep searches, none of the type 
		Ia supernova has been detected  to date. Some of the strong upper limits on 
		the radio luminosities of type Ia SNe have been presented by \citet{Panagia2006} 
		which cover time scales of  a few to several days from SN constraining 
		luminosities to $L_{\nu}<10^{25}$ erg/s/Hz. Optical sky surveys have 
		become efficient not 
		only in SN detections but also in classification, thus opening up possibilities of 
		early observations of SN Ia. As an exciting outcome of this a nearby type Ia 
		SN PTF11kly/SN2011fe was detected at a age $<1$ day and was searched 
		for radio emission by VLA. The non-detection provided one of the strongest 
		constraints on the radio luminosity and circum-stellar matter density \citep{Horesh2012,
		Chomiuk2012}. 
		Such detections will be possible more often in the near future by the upcoming
		optical transient surveys.
		
		\item Transients without high energy triggers: \citet{Horiuchi2011} pointed out that
		the SNe rate measured largely from the optical observations does not match
		the cosmic massive star formation rate. They estimate that as many as half 
		the SNe are never discovered. A possible reason could be heavy obscuration due 
		to dust in the regions where SNe occur. Interestingly, a nearby RSNe SN 2008iz
		was discovered serendipitously in the galaxy M82 which reached the peak
		luminosity $\rm L_{\nu} > 10^{27}$ erg/s/Hz \citep{Brunthaler2009}. 
		Another radio transient was discovered soon afterwards in the same galaxy
		\citep{Muxlow2010}. Luminosity of the later 
		transient $\rm L_{\nu} \approx 10^{25}$ erg/s/Hz is consistent with SNe, although
		its nature is being debated. These cases lend support to the claim that not all
		SNe could be discovered through optical and some might actually be detected 
		only via radio emission. A radio sky survey is the most efficient way to uncover
		this hidden population of SNe. 
		
		\item GW sources: A new window on the Universe is set to open  
		when the merging compact binaries, involving neutron stars and black holes, 
		will be detected through the 
		emission of gravitational waves (GW). Advanced GW detectors such as ALIGO 
		and VIRGO will be able to detect such emission from sources within 
		about 300 Mpc. Their localization, however, is going to be coarse with 
		the positional uncertainty being of the order of several tens of degrees.
		Therefore, identifying these transients through their associated 
		electromagnetic signatures, having orders of magnitude better localization 
		capabilities will be of crucial importance. It has been shown by several authors,
		e.g. \citet{Nakar2011,Kamble2013}, that mergers of binary neutron stars
		as well as binary supermassive black holes could result in subsequent bright
		radio emission. A snap-shot of the entire radio sky, at a sensitivity of 0.1 mJy 
		should be able to detect from a few to a dozen such transients within
		a horizon of size similar to that of ALIGO.

	\end{enumerate}
\section{Rate Estimates}
	A radio SN of typical luminosity $\overline{L_{\nu}}$ will be detectable from as 
	far away as 
	$d_{L} = \sqrt{\overline{L_{\nu}}/4\pi F_{\nu,lim} }$ for the assumed survey sensitivity 
	$F_{\nu,lim}$. Based on the previous observations and surveys the log-normal 
	distribution of peak spectral luminosities of RSNe peaks at $\overline{L_{\nu}} 
	\approx 2\times 10^{27}$ erg/s/Hz as shown in Figure~\ref{fig:dist}. For the fiducial
	sensitivity of $F_{\nu,lim} = 0.1$ mJy per VLA pointing a typical RSN should therefore 
	be detectable out to about $\approx 125$ Mpc or redshift $\approx 0.03$. This
	corresponds to the detectable volume in the local universe $V_{detect} \approx 
	8.2 \times 10^{6}$ Mpc$^{3}$. As far as RSNe are concerned, it is clear that the VLA 
	will typically be detecting those events in the local universe. Effects of cosmological 
	redshift could therefore be ignored from the rate calculations.
		
	If $\Re$ is the volumetric rate of supernovae in the local universe, then in an all sky
	snap-shot VLA survey should be able to detect about $N_{all-sky} = 
	\Re V \Delta t$ radio SNe within the detectable volume $V$ and lasting for time 
	$\Delta t$ above the VLA sensitivity. It is clear from the past experiences, as shown 
	in Figure~\ref{fig:dist}, that most of the RSNe reach their peak brightness around 30
	days after the burst. Therefore, we will use $\Delta t \approx 30$ days.
	
	The volumetric rate of core-collapse SNe have been estimated from observations, 
	and ranges from $10^{-3} \leq \Re \leq 10^{-4}$ SNe yr$^{-1}$ Mpc$^{-3}$. 
	Using $V=V_{detect}$ and $\Delta t \approx 30$ days as described above, one gets
	$700 \leq N_{all-sky} < 70 $. In other words, about one SN per month 
	should be detectable for every $\rm 60 ~deg^{2}$ of the sky scanned.
	
\section{Survey Strategy}
	RSNe reach peak brightness in around 30 days at 5 GHz or VLA `C' band. 
	The spectral peak of the radio emission from SNe cascades to lower frequency bands with 
	time. The peak brightness of RSNe, however, remains largely steady in most cases which 
	is to be expected when a freely expanding shock wave is ploughing through 
	the circum-stellar wind.
	
	Given these characteristics of the RSNe, VLA-C band appears to be an optimum choice
	for carrying out radio sky survey for SNe and similar transients mentioned above.
	According to the reference guide for Jansky VLA capabilities for sky surveys,
	scanning rate of $SS \approx \rm ~7.2 ~deg^{2}/hr$ of the sky could be achieved at C-band 
	for the nominal sensitivity of $F_{\nu,lim}$ = 100$\mu$Jy. In order to confirm variability,
	a suitable cadence will be necessary. The natural choice for RSNe would be 30 days,
	roughly the timescale required for RSNe to reach peak brightness. In order to achieve
	a sky coverage of about 1000 $\rm deg^{2}$ with a cadence of 30 days, VLA would need 
	to invest about 3.3 hr of observation time on a daily basis. With the inclusion of 25\% 
	overhead time for calibration etc. about 4.1 hr/day would be required to achieve 
	the sky coverage of 1000 $\rm deg^{2}$.	This translates to detection of about 16 RSNe 
	per month, conservatively, assuming lower end of the SN rate.
	
	Alternatively, surveying at a lower frequency S-band has the advantage of faster or 
	equivalently, larger sky coverage. For the same sensitivity at S-band, survey speed 
	of $\rm SS=16.53 \rm~deg^{2}/hr$ is attainable which is more than twice as fast compared 
	to the C-band. This would require a little over 2 hr-per-day for the coverage of 
	$\rm 1000~deg^{2}$ with the cadence of 30 days. No serious losses would be inflicted 
	as a result of moving to lower than C-frequency band as far as supernova science 
	case is concerned. Radio spectral peak of supernovae cascades down to lower 
	frequencies with time as $\rm (\nu_{peak}/5~GHz) = (t/30~day)^{-1}$. Thus, the spectral 
	peak would appear in S-band in about $\rm \approx 45 ~days$, allowing longer cadence
	and therefore larger sky coverage for the survey. We estimate that more than 
	$\rm 3000 ~deg^{2}$ of sky coverage could be achieved at S-band, compared 
	to $\rm 1000 ~deg^{2}$ at the C-band, for 4.1 hr/day of VLA observations owing 
	to longer cadence allowed and faster survey speeds achievable.

\end{document}